\documentclass[aps,prb,twocolumn,superscriptaddress,amsmath,amssymb,longbibliography]{revtex4-2}

\usepackage{bm}
\usepackage{graphicx}
\usepackage{amsmath}
\usepackage{hyperref}
\usepackage{epsfig}

\usepackage{epstopdf}
\usepackage[ngerman, english]{babel}

\DeclareGraphicsExtensions{.pdf}

\usepackage{hyperref}
\hypersetup{%
	colorlinks=true,
	linkcolor=blue,
	urlcolor=blue,
	citecolor=blue
}

\begin{document}

\title{Positive magnetoresistance and chiral anomaly in exfoliated type-II Weyl semimetal $T_\mathrm{d}$-WTe$_{2}$}

\author{R. Adhikari}
\email{rajdeep.adhikari@jku.at}
\affiliation{Institut f\"ur Halbleiter-und-Festk\"orperphysik, Johannes Kepler University, Altenbergerstr. 69, A-4040 Linz, Austria}

\author{S. Adhikari}
\email{soma.adhikari@jku.at}
\affiliation{Institut f\"ur Halbleiter-und-Festk\"orperphysik, Johannes Kepler University, Altenbergerstr. 69, A-4040 Linz, Austria}

\author{B. Faina}
\affiliation{Institut f\"ur Halbleiter-und-Festk\"orperphysik, Johannes Kepler University, Altenbergerstr. 69, A-4040 Linz, Austria}

\author{M. Terschanski}
\affiliation{Department of Physics, TU Dortmund, Otto-Hahn-Straße 4,	44227 Dortmund, Germany}

\author{S. Bork}
\affiliation{Department of Physics, TU Dortmund, Otto-Hahn-Straße 4,	44227 Dortmund, Germany}

\author{C. Leimhofer}
\affiliation{Institut für Polymerwissenschaften, Johannes Kepler University, Altenbergerstr. 69, A-4040 Linz, Austria}

\author{M. Cinchetti}
\affiliation{Department of Physics, TU Dortmund, Otto-Hahn-Straße 4,	44227 Dortmund, Germany}

\author{A. Bonanni}
\email{alberta.bonanni@jku.at}
\affiliation{Institut f\"ur Halbleiter-und-Festk\"orperphysik, Johannes Kepler University, Altenbergerstr. 69, A-4040 Linz, Austria}

\begin{abstract}
	
Layered van der Waals semimetallic $T_\mathrm{d}$-WTe$_{2}$, exhibiting intriguing  properties which include non-saturating extreme positive magnetoresistance (MR) and tunable chiral anomaly, has emerged as model topological type-II  Weyl semimetal system. Here, $\sim$45 nm thick mechanically exfoliated flakes of $T_\mathrm{d}$-WTe$_{2}$ are studied $via$ atomic force microscopy, Raman spectroscopy, low-$T$/high-$\mu_{0}H$ magnetotransport measurements and optical reflectivity. The contribution of anisotropy of the Fermi liquid state to the origin of the large positive transverse $\mathrm{MR}_\perp$ and the signature of chiral anomaly of the type-II Weyl fermions are reported. The samples are found to be stable in air and no oxidation or degradation of the electronic properties are observed. A transverse $\mathrm{MR}_\perp$ $\sim$1200\,\% and an average carrier mobility of $5000$\, cm$^{2}$V$^{-1}$s$^{-1}$ at $T=5\,\mathrm{K}$ for an applied perpendicular field $\mu_{0}H_{\perp} = 7\,\mathrm{T}$ are established. The system follows a Fermi liquid model for $T\leq50\,\mathrm{K}$ and the anisotropy of the Fermi surface is concluded to be at the origin of the observed positive MR. The anisotropy of the electronic behaviour is also confirmed by optical reflectivity measurements. The relative orientation of the crystal axes and of the applied electric and magnetic fields is proven to give rise to the observed chiral anomaly in the in-plane magnetotransport.

\end{abstract}

\date{\today}

\maketitle

\section*{Introduction}

The presence of accidental 2-fold degeneracies in the electronic band structures of solids leads to linear energy dispersions in the vicinity of the energy-degenerate points or nodes \cite{Herring:1937_PhysRev,Armitage:2018_RMP,Hasan:2017_Ann.Rev.CMP}. The emerging low energy excitations near these degenerate points follow a photon-like linear dispersion and can be described by the Weyl equation \cite{Weyl:1929_ZP,Pal:2011_AJP}. The signatures of such low energy Weyl fermion-like excitations in condensed matter systems were recently observed in bulk NbAs and TaAs crystals \cite{Xu:2015_NatPhys,Yang:2015_NatPhys,Lv:2015_NatPhys,Xu:2015_NatPhys}. These materials  are symmetry-protected topological states of matter \cite{Ando:2013_JPSJ,Ando:2015_Ann.Rev.,Hasan:2010_RMP,Armitage:2018_RMP,Bernevig:2018_JPSJ,He:2018_Ann.Rev,Sessi:2016_Science,Adhikari:2019_PRB} and are characterized by conduction and valence bands which join with a linear dispersion around a pair of Weyl nodes \cite{Wan:2011_PRB,Armitage:2018_RMP,Ilan:2020_NatRevPhys}. Like in high energy physics, in low energy condensed matter systems, the Dirac, Weyl and Majorana fermions constitute the family of elementary fermions \cite{Dirac:1928_PRSA,Majorana:1937_NC,Pal:2011_AJP} and are an appealing workbenck for fault-tolerant quantum technologies \cite{Nayak:2008_RMP,Alicea:2012_RPP,Goerbig:2017_Book}. The Weyl semimetals (WSM) are topological semimetals hosting Weyl quasiparticles (WQP) \cite{Hasan:2017_Ann.Rev.CMP,Yan:2017_Ann.Rev.CMP,Armitage:2018_RMP}. The WQPs are massless spin-1/2 fermions, but their dispersion resembles the one of photons, due to the effective relativistic symmetry and the gapless Weyl nodes. The momenta $k$ of the WQPs are projected either parallel or antiparallel to their spins and are distinguished through the quantum number chirality $\chi$. The quantum expectation value of the chiral current in WSM is not conserved \cite{Ilan:2020_NatRevPhys}, leading to non-conservation of the chiral current, also known as chiral anomaly or Adler-Bell-Jakiw (ABJ) anomaly \cite{Nielsen:1983_PhysLett,Lv:2017_PRL,Zhang:2016_NatCommun}. In condensed matter physics, the WSM are either of type-I or of type-II. The type-II WSM (WSM-II) are known to break the Lorentz symmetry, which is - in contrast - conserved in the type-I WSMs \cite{Armitage:2018_RMP}. The violation of the Lorentz symmetry leads to a tilted Weyl cone in the momentum space and makes these materials compelling for the study of exotic Lorentz violation theories that are beyond the Standard Model \cite{Hasan:2017_Ann.Rev.CMP}. The tilted Weyl cone combined with the broken chiral symmetry results in the onset of quantum mechanical and topological mechanisms, such as intrinsic anomalous Hall effects \cite{Kang:2019_NatMater}, Klein tunneling \cite{Yesilyurt:2016_SciRep}, Landau level collapse \cite{Arjona:2017_PRB} and ABJ anomaly \cite{Nielsen:1983_PhysLett,Adler:1969_PhysRev,Bell:1969_NCs}, both in WSM-I and in WSM-II.

\begin{figure*}[htbp]
	\centering
	\includegraphics[scale=1.2]{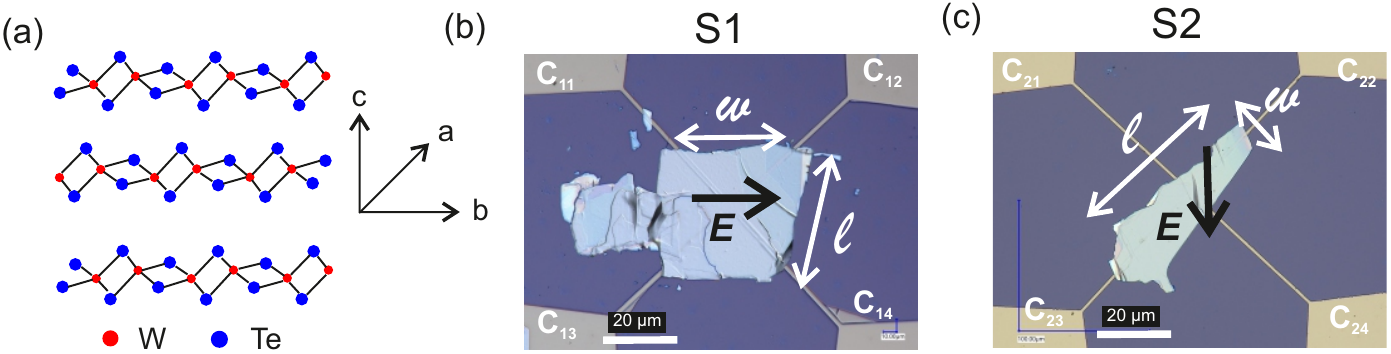}
	\caption{\label{fig:Device} (a) Schematic illustration of the crystal structure of $T_\mathrm{d}$-WTe$_{2}$ showing the directions of the $a$-, $b-$, and $c-$axes. Optical image of the WTe$_{2}$ samples S1 (b) and S2 (c), respectively. }	
	\label{fig:fig1}
\end{figure*}

The presence of tilted Weyl cones and of low energy excitations violating the Lorentz invariance in the vicinity of the Weyl points were predicted for the electronic band structure of WTe$_{2}$, a transition metal dichalcogenide (TMDC) semimetal with layered van der Waals structure \cite{Soluyanov:2015_Nature}. In addition to the presence of the Fermi arc at the Fermi surface \cite{Bruno:2016_PRB,Wu:2017_PRB,Das:2019_ElectronStruct}, anisotropic transport properties including the anisotropic ABJ anomaly are reported for both bulk and few layers WTe$_{2}$ flakes \cite{Ong:2020_arxiv,Hosur:2013_CRP,Lv:2017_PRL,Wang:2019_NanoLett,Wang:2016_PRB,Wang:2020_NatCommun,Pan:2017_FrontPhys.,Pan:2018_AdvPhysX,Liu:2017_2D,Li:2017_NatCommun,Luo:2017_Nanotechnol,Deng:2019_PRL}. The tilted Weyl cone was also observed in the TMDC semimetal MoTe$_{2}$ \cite{Liang:2019_AIPAdv,Deng:2016_NatPhys} and in LaAlGe \cite{Xu:2017_SciAdv}. Another characteristic feature of WTe$_{2}$ is the extreme positive transverse magnetoresistance $\mathrm{MR}_\perp$, reaching values as high as $10^{5}\%$ for magnetic fields $\sim$9\,T applied parallel to the $c-$axis of bulk or thin flakes of WTe$_{2}$ in the Fermi liquid phase  at $T\sim$100 mK \cite{Ali:2014_Nature,Ali:2015_EPL,Thoutam:2015_PRL}. In addition, evidence of topologically protected conducting edge states \cite{Kononov:2020_NanoLett} and flat-band superconductivity in close proximity to Pd \cite{Kononov:2020_arxiv} and Nb \cite{Huang:2020_NSR} was demonstrated in bulk and few layers WTe$_{2}$ flakes, while quantum spin Hall states  are shown to exist in mechanically exfoliated monolayer WTe$_{2}$ \cite{Shi:2019_SciAdv}. Most of the reported literature on $T_\mathrm{d}$-WTe$_{2}$ concerns bulk crystals or mechanically exfoliated flakes. While the bulk crystals are chemically stable \cite{Ali:2014_Nature}, the ultrathin exfoliated flakes of $T_\mathrm{d}$-WTe$_{2}$ are found to be prone to oxidation and require inert ambient for fabrication and sample processing \cite{Kononov:2020_NanoLett}. In particular, it is not evident whether the crystal exfoliates along a preferred direction corresponding to the $a-$ or $b-$axes \cite{Zhang:2016_NatCommun}. In addition, apart from the observed extreme positive $\mathrm{MR}_\perp$, weak-antilocalization (WAL) and negative longitudinal magnetoresistance $\mathrm{MR}_\parallel$ with chiral anomaly are reported in exfoliated flakes of $T_\mathrm{d}$-WTe$_{2}$ \cite{Zhang:2016_NatCommun}.

\begin{figure*}[htbp]
	\centering
	\includegraphics[scale=1.25]{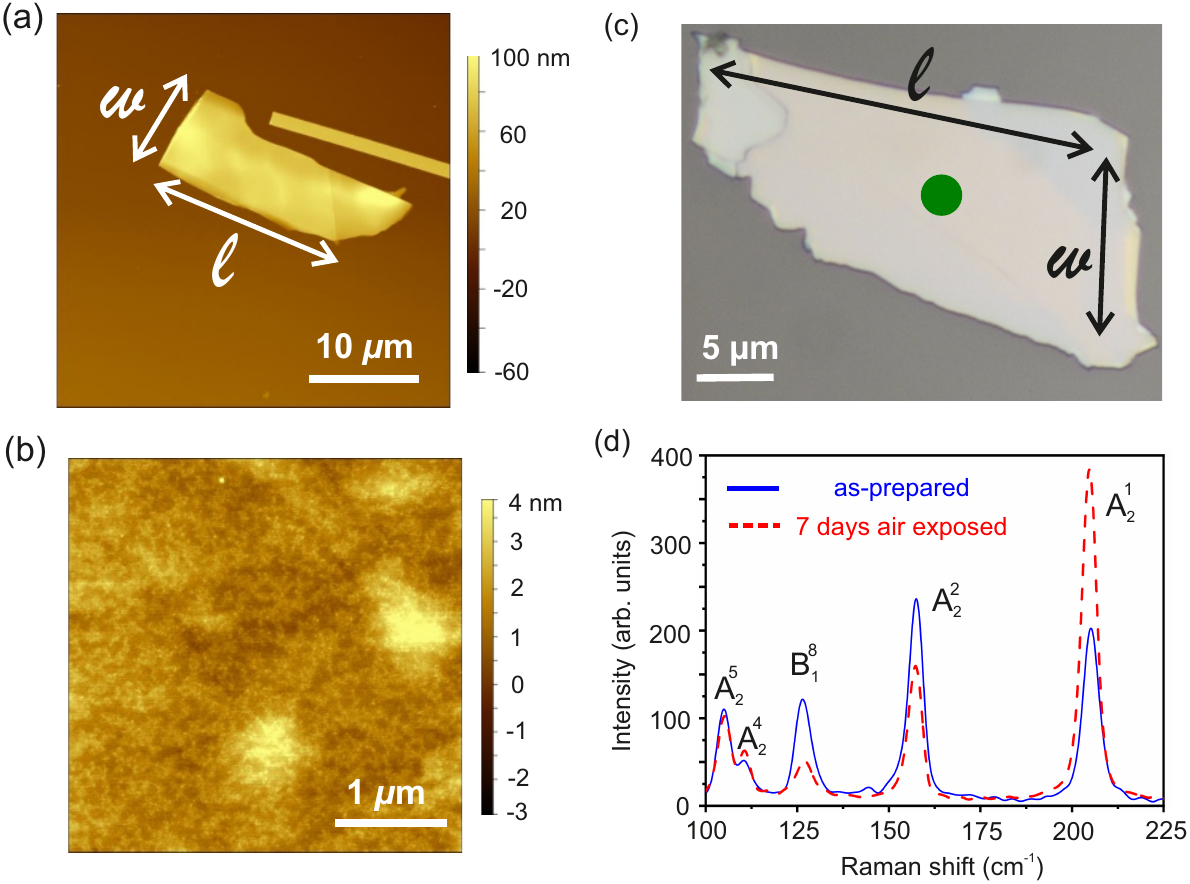}
	\caption{\label{fig:AOR} (a) AFM image of an exfoliated WTe$_{2}$ flake dry transferred onto the SiO$_{2}$/$p^{++}$-Si substrate. (b) Surface morphology recorded for a $\left(3.5\times3.5\right)\,\mu\mathrm{m}^{2}$ AFM scan area of the transferred WTe$_{2}$ flake. (c) Optical microscopic image of  45 nm thick WTe$_{2}$ flake used to measure Raman spectroscopy. The dot indicates the position of the laser spot during the Raman measurements. (d) Raman spectra measured on the specimen in (c) as-prepared and after seven days of exposure to air ambient.}	
	\label{fig:fig2}
\end{figure*}

Here, the chemical stability, Fermi liquid behavior with large positive $\mathrm{MR}_\perp$ up to $\sim$1200\% and average carrier mobility $\mu_\mathrm{av}$ $\sim$$5000$\, cm$^{2}$V$^{-1}$s$^{-1}$ of mechanically exfoliated $\sim$45\,nm thin flakes of $T_\mathrm{d}$-WTe$_{2}$ are studied $via$ atomic force microscopy (AFM), Raman spectroscopy, optical reflectivity and low-$T$/high-$\mu_{0}H$ magnetotransport measurements.

 \section*{Sample fabrication}

The WTe$_{2}$ flakes are fabricated $via$ mechanical exfoliation from a bulk (mother) crystal obtained from hqgraphene. Micromechanical cleavage is repeated several times using fresh Nitto tapes until $\sim$45 nm thin flakes are obtained. Upon exfoliation, the ﬂakes are transferred onto Gelpack Grade 4 viscoelastic polydimethylsiloxane (PDMS) stamps with a rigid metallic support. The PDMS used in this work is a Gelpack Grade 4 viscoelastic film with a rigid metallic support. The mechanical exfoliation process results in an ensemble of flakes with diverse sizes, geometries and thicknesses, distributed over the PDMS stamp. The exfoliated flakes on the PDMS are then analysed for thickness and number of layers using a high resolution Keyence VHX-7000 optical microscope operated in transmission mode. Flakes of uniform thickness are then transferred onto prepatterned SiO$_{2}$/$p^{++}$ Si substrates with markers and metal contact pads. The flakes transferred onto the pristine substrates are used for AFM, Raman spectroscopy and optical measurements, while the samples fabricated on the prepatterned substrates are employed to study the electronic properties of the exfoliated WTe$_{2}$ flakes by low-$T$/high-$\mu_{0}H$ magnetotransport measurements.

The $\left(1\times1\right)$ cm$^{2}$ SiO$_{2}$/$p^{++}$ Si substrates with a SiO$_{2}$ thickness of 90 nm are spin coated with S1805 positive photoresist followed by soft baking at 90$^{\circ}$C. A Süss Mask aligner photolithography system is employed to expose the S1805 coated substrates to an ultraviolet mercury lamp through a window mask. The substrates are developed using a conventional MS-519 photolithography developer and subsequently transferred into a sputtering chamber for the deposition of 10 nm thick Pt contacts. The metallic contact pads in van der Pauw 4-probe geometry are then obtained by rinsing away the photoresist with warm acetone for 15 seconds in an ultrasonic bath.

The exfoliated WTe$_{2}$ flakes are transferred onto both pristine and prepatterned substrates using an indigeneously developed viscoelastic dry transfer system \cite{Grundlinger:2018_Master}. Gold wires with a diameter of 25 $\mu$m are bonded to the Pt contact pads using In as conducting adhesive.  It is worth noting, that the mechanical exfoliation and the viscoelastic dry transfer are carried out in air ambient. The stability of the samples in ambient conditions is discussed in the next section. This is in contrast to what generally reported in literature, where the fabrication and transfer of the flakes takes place in vacuum or in inert atmosphere \cite{Wang:2019_NanoLett,Kononov:2020_NanoLett}. 

A schematic representation of the semimetallic bulk crystal structure of $T_\mathrm{d}$-WTe$_{2}$ highlighting the $a-$, $b-$ and $c-$ directions of the space group $C^7_{2v}$ ($Pmn2_{1}$) distorted orthorhombic basis, is shown in Fig.~\ref{fig:fig1}(a). Each unit cell is composed of two W and four Te atoms and the W-Te bond lengths vary between $2.7\,\text{{\AA}}$ and $2.8\,\text{{\AA}}$ \cite{Ali:2014_Nature,Augustin:2000_PRB}. The bulk WTe$_{2}$ exhibits a $T_\mathrm{d}$ stacking order in which the atoms in the upper layer are rotated by 180$^\circ$ w.r.t. the atoms in the lower layer, as sketched in Fig.~\ref{fig:fig1}(a). In the $T_\mathrm{d}$-WTe$_{2}$, the $a$-axis is populated by the W-chain while the axis $b$ is orthogonal to it. The $c-$axis is perpendicular to the $ab-$ plane. Two exemplary samples are considered, namely: 

(i) Sample S1: in the van der Pauw geometry the excitation current $I_\mathrm{ac}$ is applied between the contacts $\mathrm{C}_{11}$ and $\mathrm{C}_{12}$ while the resulting voltage $V_\mathrm{xx}$ is measured across $\mathrm{C}_{13}$ and $\mathrm{C}_{14}$, so that the electric field $E$ due to $I_\mathrm{ac}$ is aligned exactly along $w$, as visualized in the optical microscopy image in Fig.~\ref{fig:fig1}(b);
\\[8pt]
(ii) Sample S2: in the van der Pauw geometry $I_\mathrm{ac}$ is applied between the contacts $\mathrm{C}_{21}$ and $\mathrm{C}_{22}$ while $V_\mathrm{xx}$ is measured across $\mathrm{C}_{23}$ and $\mathrm{C}_{24}$, so that $E$ is slightly misaligned with respect to $w$, as evidenced in the optical microscopy image given in Fig.~\ref{fig:fig1}(c). 
\\[8pt]
The length and width of the rectangle-like flakes are indicated by $l$ and $w$, respectively. In the absence of a conclusive evidence for the precise orientations of the $a-$ and $b-$axes, the geometry of the studied flakes is described by $l$ and $w$, while the $c-$axis is the one perpendicular to the plane of the flakes.

 \section*{Results and Discussion}
 
  \subsection*{Atomic force microscopy and Raman spectroscopy}
  
The surface morphology and the thickness of the WTe$_{2}$ flakes are measured using a VEECO Dimension 3100 AFM system. The AFM image of a WTe$_{2}$ flake transferred onto a SiO$_{2}$/$p^{++}$ Si substrate is provided in Fig.~\ref{fig:fig2}(a). The height profile of the flake is analysed using the Gwiddyon data analysis software and a thickness of 45 nm is determined. The surface morphology of the flake is shown in Fig.~\ref{fig:fig2}(b) and a root mean square surface roughness of $\sim$0.45 nm is estimated.

The chemical stability and structural phase of the WTe$_{2}$ flakes are studied using Raman spectroscopy carried out in a WIRec Alpha 300 R-Raman-System with a double frequency Nd:YAG laser of wavelength 532 nm. The objective allows a laser beam spot diameter of $\sim$2\,$\mu$m. The  samples are positioned on a XY-translation stage and a camera system enables guiding the sample in the laser spot. A total of 33 Raman vibrations are predicted by group theory \cite{Jiang:2016_SciRep} and the irreducible representation of the optical phonons at the $\Gamma$ point of the Brilloiun zone of the bulk  $T_\mathrm{d}-$WTe$_{2}$ is given by:

\begin{equation}
	\Gamma_\mathrm{bulk} =11A_{1}+6A_{2}+11B_{1}+5B_{2}
\end{equation} 
where $A_{1}$, $A_{2}$, $B_{1}$ and $B_{2}$ are Raman active phonon modes. In this work, the Raman modes have been excited along the $c-$axis of the $T_\mathrm{d}-$WTe$_{2}$ crystal $i.e.$ the laser beam is directed perpendicular to the plane of the WTe$_{2}$ flake and of the substrate. Since the Raman excitations reported here are unpolarized, neither the incident, nor the scattered electric field vectors $\vec{e_{\mathrm{i}}}$ and $\vec{e_{\mathrm{s}}}$ are aligned along the $a-$ or $c-$axes. An optical microscopy image of the flake studied by Raman spectroscopy is shown in Fig.~\ref{fig:fig2}(c), while the room temperature Raman spectra recorded for the as-prepared and air aged WTe$_{2}$ flake are given in Fig.~\ref{fig:fig2}(d). A total of five Raman active modes with peaks centered at 105.0 cm$^{-1}$, 110.5 cm$^{-1}$, 126.5 cm$^{-1}$, 157.2 cm$^{-1}$, and 205.3 cm$^{-1}$ have been recorded. The Raman active modes are labelled as $A_{\mathrm{q}}^{\mathrm{p}}$ and $B_{\mathrm{q}}^{\mathrm{p}}$, where $\mathrm{q}=\{1, 2\}$ and $\mathrm{p}\in{\mathbb{Z}}$ to uniquely identify the Raman mode. Here, the five recorded Raman peaks are assigned to the $A^{5}_{2}$, $A^{4}_{2}$, $B^{8}_{1}$, $A^{2}_{2}$, and $A^{1}_{2}$ allowed Raman active modes \cite{Jiang:2016_SciRep,Augustin:2000_PRB}. Upon comparison with the calculated resonances, it is found that the experimental $A^{4}_{2}$ is blue shifted, while the other four Raman active modes are red shifted due to the stress built up when the flake is transferred onto the rigid substrate. Upon measurement, the sample has been exposed to air for seven days and then Raman spectra have been recorded. The Raman spectrum for the air aged sample, represented by the  dashed curve in Fig.~\ref{fig:fig2}(d) is found to show the allowed Raman modes of the optical phonons as recorded for the pristine flake. No peaks related to oxides of Te or W are detected, confirming the chemical stability of the WTe$_{2}$ flakes.         
\begin{figure}[htbp]
	\centering
	\includegraphics[scale=0.70]{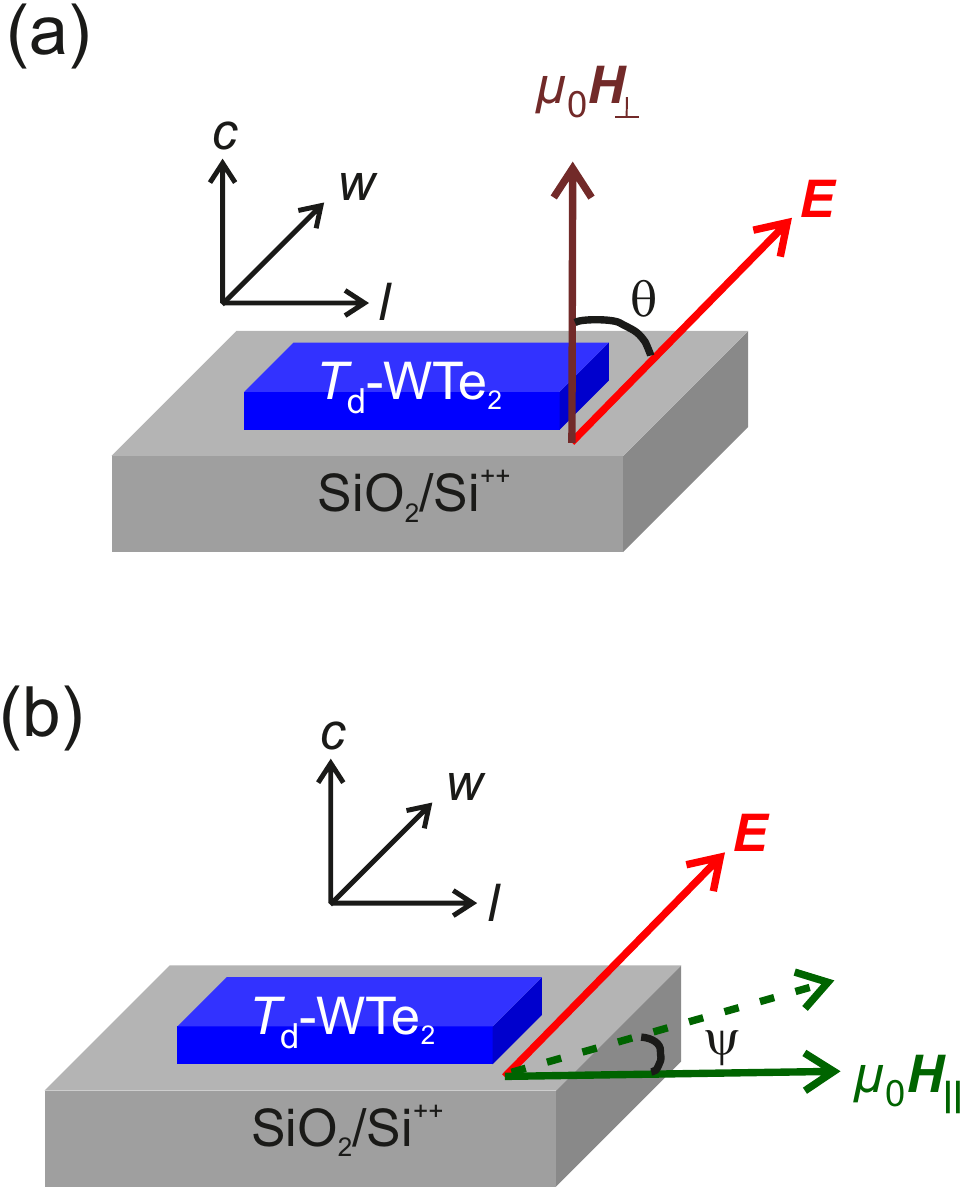}
	\caption{\label{fig:CON} Schematic illustration of the relative orientations of $E$, $\mu_{0}H$, $l$ and $w$ relevant for the measurements of (a) $\mathrm{MR}_\perp$ and (b) $\mathrm{MR}_\parallel$. }	
	\label{fig:fig3}
\end{figure}
 
   \subsection*{Out-of-plane magnetotransport} 
    
\begin{figure*}[htbp]
	\centering
	\includegraphics[scale=2.75]{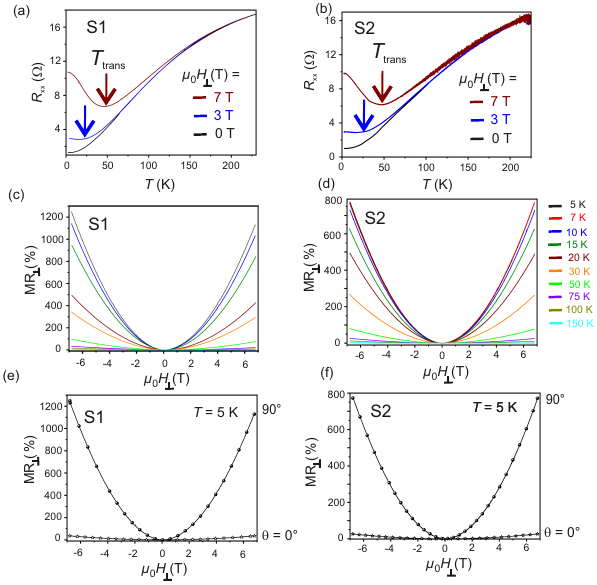}
	\caption{\label{fig:MR} (a) and (b): $R_\mathrm{xx}$ as a function of $T$ at $\mu_{0}H_\perp=0\,\mathrm{T}$, $\mu_{0}H_\perp=3\,\mathrm{T}$ and $\mu_{0}H_\perp=7\,\mathrm{T}$ for S1 and S2, respectively. (c) and (d): $\mathrm{MR}_\perp$ as a function of $\mu_{0}H_\perp$ measured in the range $5\,\mathrm{K}\leq{T}\leq150\,\mathrm{K}$ for S1 and S2, respectively. (e) and (f): $\mathrm{MR}_\perp$ as a function of $\mu_{0}H$ at $\theta=0^\circ$ and $\theta=90^\circ$ at $T=5\,\mathrm{K}$ for S1 and S2, respectively.}	
	\label{fig:fig4}
\end{figure*}

Low-$T$/high-$\mu_{0}H$ magnetotransport measurements are carried out on S1 and S2 in a Janis Super Variable Temperature 7TM-SVM cryostat equipped with a 7\,T superconducting magnet. Prior to the measurements, the Ohmic nature of the Pt contacts to the WTe$_{2}$ flakes is confirmed by the linear $I-V$ characteristics measured with a high precision Keithley 4200 SCS dc source-measure unit (SMU). The longitudinal resistance $R_\mathrm{xx}$ as a function of $T$ and $\mu_{0}H$ has been measured by employing a lock-in amplifier (LIA) ac technique. The $I_\mathrm{ac}$ is sourced from a Standord Research SR830 LIA, while the generated voltage $V_\mathrm{xx}$ is measured in a phase locked mode as a function of $T$ and $\mu_{0}H$. The applied current is limited to 10 $\mu$A to minimize Joule heating and subsequent thermogalvanic effects due to the constraints imposed by the low dimensionality of the samples. The lock-in expand function is employed to enhance the sensitivity of the LIA. All measurements have been performed at a frequency of 127 Hz. The chosen reference axes for the applied $\mu_{0}H$, for $E$ due to $I_\mathrm{ac}$, and for the $l$ and $w$ dimensions of the specimens, identified to characterize the transverse magnetoresistance $\mathrm{MR}_\perp$ and the longitudinal magnetoresistance $\mathrm{MR}_\parallel$, are shown in Fig.~\ref{fig:fig3}(a) and Fig.~\ref{fig:fig3}(b) for S1 and S2, respectively. In Fig.~\ref{fig:fig3}(a), $\theta$ is the angle between $\mu_{0}H_\perp$ applied along the $c-$axis and the $ab-$plane, while $E$ is oriented along $w$. The $\mu_{0}H_\perp$ has been recorded for $\theta=0^\circ$ and $90^\circ$. The $\mathrm{MR}_\parallel$ for S1 and S2 are measured by applying an in-plane magnetic field $\mu_{0}H_\parallel$ at an angle $\psi$ w.r.t. $E$, while $E$ is always applied along $w$. Thus, there are two possible configurations of the relative orientations of $E$, $\mu_{0}H_\parallel$ w.r.t. $w$, namely: (i) $\mu_{0}H_{\parallel}\parallel\left({E}\parallel{w}\right)$ and (ii) $\mu_{0}H_{\parallel}\perp\left({E}\parallel{w}\right)$.  
  
\begin{figure*}[htbp]
	\centering
	\includegraphics[scale=2.0]{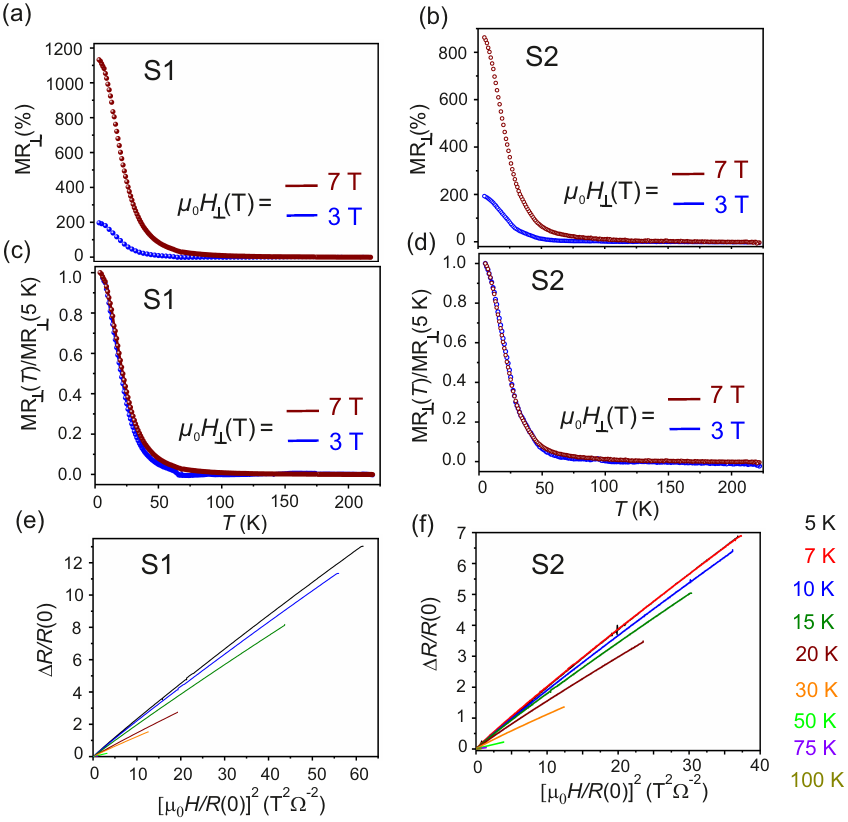}
	\caption{\label{fig:MRT} (a) and (b): FC transverse $\mathrm{MR}$ as a function of $T$ recorded for $\mu_{0}H_{\perp}=3\,\mathrm{T}$ and $7\,\mathrm{T}$ for samples S1 and S2, respectively. (c) and (d): estimated normalized $\mathrm{MR}_\perp$ defined as the ratio of $\mathrm{MR}_{\perp}(T)$ to  $\mathrm{MR}_{\perp}(5\,\mathrm{K})$ recorded by applying $\mu_{0}H_\perp=3\,\mathrm{T}$ and $\mu_{0}H_\perp=7\,\mathrm{T}$ for S1 and S2, respectively. (e) and (f): calculated Kohler's plots of S1 and S2 in the range $5\,\mathrm{K}\leq{T}\leq100\,\mathrm{K}$.}	
	\label{fig:fig5}
\end{figure*}

\begin{figure*}[htbp]
	\centering
	\includegraphics[scale=1.75]{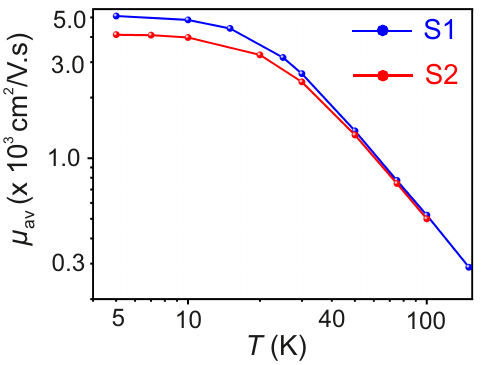}
	\caption{\label{fig:MOB} $\mu_\mathrm{av}$ as a function of $T$ for samples S1 and S2.}	
	\label{fig:fig6}
\end{figure*}

The evolution of $R_\mathrm{xx}$ as a function of $T$ in the interval $5\,\mathrm{K}\leq{T}\leq225\,\mathrm{K}$ for S1 and for S2 is given in Fig.~\ref{fig:fig4}(a) and Fig.~\ref{fig:fig4}(b), respectively. The $R_\mathrm{xx}$--$T$ behavior is studied while the samples are cooled down, both for $\mu_{0}H_{\perp}=0$ and for $\mu_{0}H_{\perp}\neq{0}$ and the measurements are referred to as zero field cooled (ZFC) and  field cooled (FC), respectively. The FC $R_\mathrm{xx}$ are measured by applying $\mu_{0}H_{\perp}=$\, 3\,T and 7\,T, as shown in Fig.~\ref{fig:fig4}(a) and Fig.~\ref{fig:fig4}(b). The monotonous decrease of $R_\mathrm{xx}$ with decreasing $T$ for $\mu_{0}H_{\perp}=0$ is a signature of the metallic behavior of the exfoliated $T_\mathrm{d}$-WTe$_{2}$ flakes. For $\mu_{0}H_{\perp}\geq{\mathrm{3\,T}}$, $R_{\mathrm{xx}}$ essentially follows the ZFC behavior down to $T_{\mathrm{Trans}}$. At $T=T_{\mathrm{Trans}}$ -- indicated by the arrows ($\downarrow$) in Fig.~\ref{fig:fig4}(a) and Fig.~\ref{fig:fig4}(b) -- a transition in the electronic phase of the flakes from metallic to insulating is observed for both S1 and S2. This behavior is consistent with the one previously observed for bulk and thin $T_\mathrm{d}$-WTe$_{2}$ layers \cite{Ali:2014_Nature,Luo:2017_Nanotechnol,Thoutam:2015_PRL,Zhang:2017_nanoLett,Zhao:2015_PRB}. It is also noted, that the magnitude of the $\mu_{0}H_{\perp}$- induced change in $R_\mathrm{xx}$ increases with decreasing $T$ and increasing $\mu_{0}H_{\perp}$. The magnetoresistance, defined as:

\begin{equation}
\mathrm{MR}=\frac{R_{\mathrm{xx}}(\mu _{0}H)-R_{\mathrm{xx}}(0)}{R_{\mathrm{xx}}(0)}\times{100\%}
\end{equation}

is a fingerprint of the microscopic physical mechanism governing the electronic properties of any trivial or non-trivial electronic system. Here, $R_{\mathrm{xx}}(\mu _{0}H)$ and $R_{\mathrm{xx}}(0)$ are the resistances of the system under an applied field $\mu_{0}H$ and in zero field, respectively. The  $\mathrm{MR}_\perp$ of the two samples S1 and S2 are estimated as a function of an applied field $\mu_{0}H_{\perp}$ at different $T$ in the range $5\,\mathrm{K}\leq{T}\leq150\,\mathrm{K}$ and are reported in Fig.~\ref{fig:fig4}(c) and Fig.~\ref{fig:fig4}(d). Large positive $\mathrm{MR}_\perp$ $\sim$1200\% and $\sim$800\% are found at $T=5\,\mathrm{K}$, under the maximum applied field $\mu _{0}H_\perp=\,7\,\mathrm{T}$ for S1 and S2, respectively. The estimated $\mathrm{MR}_\perp$ for both samples follow the power-law behavior $\mathrm{MR}\sim\mathrm{m}{H^{n}}$ \cite{Abrikosov:2017_Book,Thoutam:2015_PRL}, where $\mathrm{m}$ is a constant of proportionality and $n\in\mathbb{R}$ is the power index. The value of $n$ is estimated by numerical fitting of the $\mathrm{MR}_\perp$ behavior for both S1 and S2 as a function of $\mu_{0}H_{\perp}$ and is found to be $1.9\leq{n}\leq2.1$ in the range $5\,\mathrm{K}\leq{T}\leq150\,\mathrm{K}$. The quadratic dependence of $\mathrm{MR}_\perp$ on the applied $\mu_{0}H_{\perp}$ observed for the samples reported here is in agreement with previous reports \cite{Ali:2014_Nature,Ali:2015_EPL,Thoutam:2015_PRL,Luo:2017_Nanotechnol}. It is noted, that no Shubnikov-de Haas (SdH) oscillations have been observed in the $\mathrm{MR}_\perp$ of both samples, even up to the maximum $\mu_{0}H_\perp$. The $\mathrm{MR}_\perp$ as a function of $\mu _{0}H_\perp$ for $\theta=0^{\circ}\,\mathrm{and}\,90^{\circ}$ and $T=5\,\mathrm{K}$ are also recorded for S1 and S2 and reported in Fig.~\ref{fig:fig4}(e) and Fig.~\ref{fig:fig4}(f), respectively. An anisotropic behavior of $\mathrm{MR}_\perp$ is observed for the $T_{d}$-WTe$_{2}$ flakes studied here, in agreement with the literature \cite{Thoutam:2015_PRL,Lv:2017_PRL,Lv:2015_NatPhys,Zhao:2015_PRB}.

The evolution of $\mathrm{MR}_\perp$ as a function of $T$ for $\mu _{0}H_\perp=3\,\mathrm{T}$ and $\mu _{0}H_\perp=7\,\mathrm{T}$ is shown in Fig.~\ref{fig:fig5}(a) and Fig.~\ref{fig:fig5}(b) for S1 and S2, respectively. The positive $\mathrm{MR}_\perp$ sets in for $T\leq75\,\mathrm{K}$ and for a critical field $\mu_{0}H_\mathrm{c}\geq3\,\mathrm{T}$, in accord with the $R_\mathrm{xx}$-$T$ behavior previously discussed. A relevant feature of the observed large $\mathrm{MR}_\perp$, is the presence of a magnetic field-dependent critical turn-on temperature $T_\mathrm{Trans}$ for $\mu_{0}H_{\perp}\geq3\,\mathrm{T}$. However, such critical temperature is absent when the samples are cooled down in the presence of the field $\mu_{0}H_{\parallel}$. The transition from metallic to insulating state has been observed in several other material systems exhibiting colossal positive $\mathrm{MR}$, where the extreme magnitude of the MR is attributed to a magnetic field-driven metal-to-insulator (MIT) transition \cite{Leahy:2018_PNAS,Wang:2014_SciRep,Takatsu:2013_PRL}, due to a field-induced excitonic gap in the linear spectrum of the Coulomb interacting quasiparticles, leading to an excitonic insulator phase \cite{Thoutam:2015_PRL,Khveshchenko:2001_PRL}. The excitonic gap is expected to follow the relation: 
$\Delta_{T}(\mu_{0}H-\mu_{0}H_\mathrm{c})\rightarrow\left[\mu_{0}H-\mu_{0}H_\mathrm{c}(T)\right]^\frac{1}{2}$, where $\mu_{0}H_\mathrm{c}$ is the threshold magnetic field and the dependence of the excitonic gap on the magnetic field is characteristic of a second order phase transition.

The normalized $\mathrm{MR}_\perp$ for S1 and S2, defined as the ratio between $\mathrm{MR}_\perp$ measured at any $T$ ($\mathrm{MR}_{\perp}(T)$) and $\mathrm{MR}_\perp$ at $T=5\,\mathrm{K}$ ($\mathrm{MR}_{\perp}(5\,\mathrm{K})$), are plotted as a function of $T$ for $\mu_{0}H_{\perp}=3\,\mathrm{T}$ and $\mu_{0}H_{\perp}=7\,\mathrm{T}$ in Fig.~\ref{fig:fig5}(c) and Fig.~\ref{fig:fig5}(d), respectively. It is observed, that the $\mathrm{MR}_\perp$ have the same $T$-dependence for both S1 and S2 WTe$_{2}$ flakes, as inferred from the collapse of the two curves. This behavior of the normalized $\mathrm{MR}_\perp$ is inconsistent with the existence of a $\mu_{0}H$-dependent excitonic gap \cite{Thoutam:2015_PRL,Khveshchenko:2001_PRL,Zhao:2015_PRB}. It is, thus, concluded that the origin of the MIT observed here in the $R_\mathrm{xx}$--$T$ behavior for $\mu_{0}H_{\perp}\geq3\,T$, lies in the evolution of the electronic structure of $T_\mathrm{d}$-WTe$_{2}$. From angle-dependent photoemission spectroscopic studies on bulk $T_\mathrm{d}$-WTe$_{2}$, it was shown that the presence of minute electron and hole pockets of equal size at low $T$ is responsible for the remarkably large positive $\mathrm{MR}_\perp$, due to a $T$-dependent charge compensation mechanism \cite{Das:2019_ElectronStruct,Pletokosic:2014_PRL}. The anisotropic property of the Fermi surface of $T_\mathrm{d}$-WTe$_{2}$ is reflected in an anisotropic $\mathrm{MR}_\perp$ as a function of the direction of $\mu_{0}H$ defined by $\theta$, which is also observed in the $T_\mathrm{d}$-WTe$_{2}$ flakes measured in this work, as evidenced in Fig.~\ref{fig:fig5}(e) and Fig.~\ref{fig:fig5}(f). Similar results are obtained when a 45 nm thick $T_\mathrm{d}$-WTe$_{2}$ flake is contacted with Au (instead of Pt) and the results are shown in Fig. S4 of the Supplementary Material. Therefore, it can be concluded, that the transverse magnetotransport properties of mechanically exfoliated $T_\mathrm{d}$-WTe$_{2}$ are independent of the metal employed to contact the semimetallic flakes and also of the exposure of the flakes to air ambient.

\begin{figure*}[htbp]
	\centering
	\includegraphics[scale=2.0]{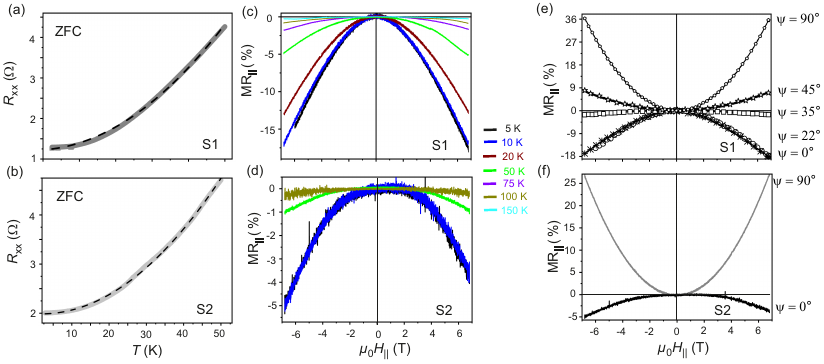}
	\caption{\label{fig:CHI}  (a) and (b): $R_\mathrm{xx}$ as a function of $T$ with $\mu_{0}H_\perp=0\,\mathrm{T}$ for S1 and S2, respectively. (c) and (d):  $\mathrm{MR}_{\parallel}$ recorded as a function of $\mu_{0}H_\parallel$ in the range $5\,\mathrm{K}\leq{T}\leq100\,\mathrm{K}$ for S1 and S2, respectively.  (e) and (f): $\mathrm{MR}_{\parallel}$ as a function of the azimuthal angle $\psi$ at $T=5\,\mathrm{K}$ for S1 and S2, respectively.}	
	\label{fig:fig7}
\end{figure*}

The presence of the electron and hole pockets in the elctronic bands of $T_\mathrm{d}$-WTe$_{2}$  is probed by Kohler's analysis of the $R_\mathrm{xx}$--$\mu_{0}H_\perp$ curves at different $T$. According to the Kohler's theory, the change in the isothermal longitudinal resistance $R_\mathrm{xx}$ for a conventional metal in an applied field $\mu_{0}H$ obeys the functional relation:
\begin{equation}
\frac{\Delta{R} _\mathrm{xx}}{R(0)}=F\Big(\frac{\mu_{0}H}{R(0)}\Big)
\end{equation}

where $R(0)$ is the ZFC resistance at $T$. The Kohler's behavior is due to the fact, that conventional metals host a single kind of charge carriers. In a weak field limit, the $\mathrm{MR}$ of most metals follows a  quadratic dependence on $\mu_{0}H$, \textit{i.e.} $\mathrm{MR}\propto\left(\alpha+\beta{\mu_{0}H^2}\right)$, with $\alpha$ and $\beta$ proportionality constants. The resistance  $R(0)\,\mathrm{is}\,\propto\frac{1}{\tau}$, where $\tau$ is the scattering time of the itinerant charge carriers in a metallic system. Therefore, a plot of $\frac{\Delta{R} _\mathrm{xx}}{R(0)}$ \textit{vs.} $\Big(\frac{H}{R(0)}\Big)^{2}$ is expected to collapse into a single $T$-independent curve, provided that the number of charge carriers, the type of charge carriers, and the electronic disorder in the system remain constant over the measured $T$ range. The Kohler's plots for S1 and S2 measured at various $T$ are reported in Fig.~\ref{fig:fig5}(e) and Fig.~\ref{fig:fig5}(f), respectively. A significant deviation is observed in the scaled transverse $\mathrm{MR}_\perp$. Due to spin dependent scattering, such deviations are common in magnetically doped topological systems \cite{Adhikari:2019_PRB}, but in a non-magnetic system such as $T_\mathrm{d}$-WTe$_{2}$ this behavior indicates that the electronic bands of the system contain both electrons and holes as charge carriers. Therefore, the formation of excitons leads to a change in the carrier density, resulting in the observed deviation from the Kohler's rule in the thin flakes of $T_\mathrm{d}$-WTe$_{2}$ as observed previously in bulk systems \cite{Pletokosic:2014_PRL,Thoutam:2015_PRL}. However, as reported by Wang $et\,al.$ \cite{Wang:2015_PRB}, bulk $T_\mathrm{d}$-WTe$_{2}$ crystal grown by chemical vapour transport follows the Kohler's rule, which is in contrast to the behavior observed here.


\begin{table*}
	\centering
	\caption{Estimated values of $\mathrm{MR}_\perp$, $\mathrm{MR}_\parallel$ and $\mu_\mathrm{av}$ at $T=5\,\mathrm{K}$ for the samples S1 and S2.}
	\begin{tabular}{|l|l|l|l|l|}
		\hline
		& $\mathrm{MR}_\perp$(\%); $T=5\,\mathrm{K}$; $\theta=90^{\circ}$ & $\mathrm{MR}_\parallel$(\%); $T=5\,\mathrm{K}$; $\psi=90^{\circ}$ & $\mathrm{MR}_\parallel$(\%); $T=5\,\mathrm{K}$; $\psi=0^{\circ}$ & $\mu_\mathrm{av}$$\left(\mathrm{cm^{2}}/\mathrm
		V.s\right)$; $T=5\,\mathrm{K}$ \\ \hline
		S1 & 1200 & 36 & -18 & 5000 \\ \hline
		S2 & 800 & 27 & -5 & 4100 \\ \hline
	\end{tabular}
\end{table*}

The values of average carrier mobility $\mu_\mathrm{av}$, $i.e.$ the mean value of the electron and hole mobilities, are calculated for both samples considered here, by fitting $\mathrm{MR}_\perp$ with the Lorentz law \cite{Abrikosov:2017_Book,Ali:2015_EPL} according to the relation:

\begin{equation}
	\frac{\Delta{R}}{R\left(0\right)}=\left[1+\left(\mu_\mathrm{av}\mu_{0}H\right)^{2}\right]
\end{equation}

The estimated $\mu_\mathrm{av}$ for S1 and S2 as a function of $T$ are presented in Fig.~\ref{fig:fig6}. Due to the dominant electron-phonon correlation, a monotonic decrease of $\mu_\mathrm{av}$ is observed for $T\,\textgreater\,50\,\mathrm{K}$, both in S1 and S2. However, for $T\leq50\,\mathrm{K}$, the estimated values of $\mu_\mathrm{av}$ with decreasing $T$ exhibit a plateau in the logarithmic scale with an estimated $\mu_\mathrm{av}\simeq{5000}$\, cm$^{2}$V$^{-1}$s$^{-1}$ and $\simeq{4000}$\, cm$^{2}$V$^{-1}$s$^{-1}$ at $T=5\,\mathrm{K}$ for S1 and S2, respectively. The value of $\mu_\mathrm{av}$ obtained in this work is higher than the one reported in literature for ultra-thin $\sim$9 nm flakes \cite{Luo:2017_Nanotechnol} measured at $T = 5\,\mathrm{K}$, but is comparable to the one reported for bulk crystals of $T_\mathrm{d}$-WTe$_{2}$ grown with the Te-flux method \cite{Luo:2015_APL}. The high carrier mobility in semimetallic $T_\mathrm{d}$-WTe$_{2}$ at low $T$ points at a deviation from the electron-phonon coupling-dominated carrier transport.

A deviation from linear behavior originating from the electron-phonon coupling at  $T\geq50\,\mathrm{K}$  is observed in the zero-field cooled $R_\mathrm{xx}$ as a function of $T$ for both S1 and S2, as previously reported in Fig.~\ref{fig:fig4}(a) and Fig.~\ref{fig:fig4}(b), respectively. The behavior of $R_\mathrm{xx}$ as a function of $T$ in the range $5\,\mathrm{K}\leq{T}\leq50\,\mathrm{K}$ is presented in Figs.~\ref{fig:fig7}(a) and \ref{fig:fig7}(b) for S1 and S2 and it follows the predictions of the Fermi liquid theory \cite{Abrikosov:2017_Book,Thoutam:2015_PRL,Pletokosic:2014_PRL} with $R_\mathrm{xx}=\gamma_\mathrm{0}+\gamma^\prime{T^2}$, where $\gamma_\mathrm{0}$ and $\gamma^\prime$ are proportionality constants. For $5\,\mathrm{K}\leq{T}\leq50\,\mathrm{K}$, electron-electron correlation is found to be the dominant mechanism in the Fermi liquid state of thin flakes of semimetallic $T_\mathrm{d}$-WTe$_{2}$ \cite{Thoutam:2015_PRL,Pletokosic:2014_PRL,Das:2019_ElectronStruct}. Thus, it can be concluded, that the large positive $\mathrm{MR}_\perp$ observed in exfoliated thin $T_\mathrm{d}$-WTe$_{2}$ flakes occurs in the Fermi liquid state of the system, as endorsed by the observed anisotropic behavior of $\mathrm{MR}_\perp$, which is a signature of an anisotropic Fermi surface \cite{Ali:2014_Nature,Ali:2015_EPL,Thoutam:2015_PRL,Pletokosic:2014_PRL,Das:2019_ElectronStruct}.  

\begin{figure*}[htbp]
	\centering
	\includegraphics[scale=1.5]{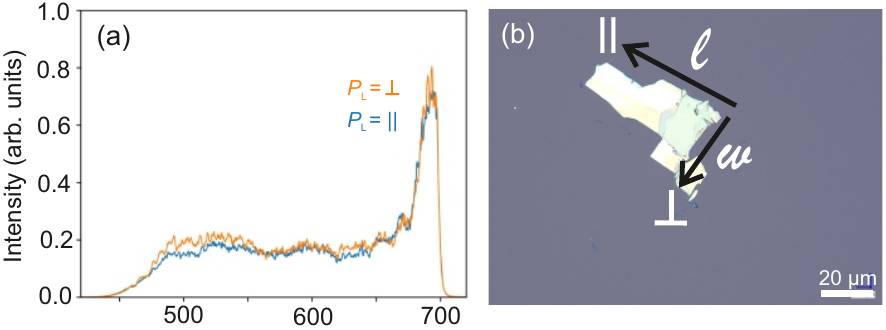}
	\caption{\label{fig:RO}  (a) Reference spectra of the supercontinuum pulses recorded for two crossed polarizations. The peak at $\sim$675 nm is associated with the seed pulse used for supercontinuum generation. (b) Optical microscopy image of the $\sim$45 nm $T_\mathrm{d}$-WTe$_{2}$ sample. The directions of the linear polarization used in the experiments ($\parallel$ and $\perp$ to the flake long axis) are also indicated in the image.}	
	\label{fig:fig8}
\end{figure*}

  \subsection*{In-plane magnetotransport}

The electronic band structure of WSM-II is characterized by the presence of an asymmetric electron dispersion responsible for the anisotropy in the electronic properties of these systems. The breaking of the Lorentz invariance and of the chiral symmetry in massless Weyl fermions under quantum fluctuations, leads to the chiral anomaly, which is observed as a negative $\mathrm{MR}_\parallel$ under the condition of $\mu_{0}\vec{H}\parallel{\vec{E}}$. In particular, for the case of $T_\mathrm{d}$-WTe$_{2}$, for $\mu_{0}\vec{H}\parallel{\vec{E}}\parallel{\vec{b}}$, the signature negative $\mathrm{MR}_\parallel$, anisotropic in the $ab$-plane of the orthorhombic lattice, is observed \cite{Zhang:2017_nanoLett,Zhang:2016_NatCommun}. In the mechanically exfoliated flakes studied here, the directions of the $a-$ and the $b-$axes are not determined $a\,priori$. As discussed above, the orientation of the studied flakes is therefore defined in terms of the dimensions $w$ and $l$. Here, $E$ is always parallel to $w$, while $\mu_{0}H_{\parallel}$ is applied at an azimuthal angle $\psi$ w.r.t. $\left(E\parallel{w}\right)$. The angle $\psi$ is varied between $0^\circ$ and $90^\circ$. The $\mathrm{MR}_\parallel$ has been estimated for both S1 and S2. The recorded $\mathrm{MR}_\parallel$ for S1 and S2 at $T=5\,\mathrm{K}$ are reported in Figs.~\ref{fig:fig7}(c) and \ref{fig:fig7}(d), respectively. For S1, $\mathrm{MR}_\parallel$ is recorded for $\psi=0^\circ,\,22^\circ,\,35^\circ,\,45^\circ\,\mathrm{and}\,90^\circ$. It is found that the negative $\mathrm{MR}_\parallel$ due to the chiral anomaly disappears for $\psi\gtrsim35^{\circ}$. For sample S2, $\mathrm{MR}_\parallel$ is collected for $\psi=0^\circ,\,\mathrm{and}\,90^\circ$ solely to show the signature of anisotropy in $\mathrm{MR}_\parallel$ and the chiral anomaly. Negative $\mathrm{MR}_\parallel$ of magnitude -18\% and -5\% for $\mu_{0}H_{\parallel}=7\,\mathrm{T}$ are observed at $T = 5\,\mathrm{K}$ and $\psi=0^\circ$ for S1 and S2, respectively. The negative $\mathrm{MR}_\parallel$ for $\mu_{0}H_{\parallel}\parallel\left({E}\parallel{w}\right)$ is a signature of  chiral anomaly. The magnitude of $\mathrm{MR}_\parallel$ decreases with increasing $T$ and vanishes at $T\geq150\,\mathrm{K}$, as shown in Figs.~\ref{fig:fig7}(c) and \ref{fig:fig7}(d) for S1 and S2, respectively.

The magnitude of the negative $\mathrm{MR}_\parallel$ is reduced on deviating from the parallel condition $i.e.$ for $\psi>0^\circ$. As depicted in Fig.~\ref{fig:fig7}(e), the magnitude of $\mathrm{MR}_\parallel$ estimated at $\psi=22^\circ$ is comparable to the one assessed for $\psi=0^\circ$. However, for $\psi=35^\circ$, a $\mathrm{MR}_{\parallel}\sim$-1\% is determined, which then reverses sign for increasing $\psi$ and a positive $\mathrm{MR}_{\parallel}$ is measured for $\psi=45^\circ$ and $\psi=90^\circ$, respectively. Therefore, it  is inferred that the assigned $w$ axis of the flakes studied here is indeed aligned along the $b-$axis of the distorted rhombohedral unit cell of $T_\mathrm{d}$-WTe$_{2}$. Further, it is also established that the topological pumping of the chiral charge current in the WSM-II $T_\mathrm{d}$-WTe$_{2}$ occurs when $\mu_{0}H_{\parallel}$ is applied at angles $\psi<35^\circ$ w.r.t. $E\parallel{w}$. For $\mu_{0}H_{\parallel}\perp\left(W\parallel{w}\right)$ $i.e.$ at $\psi=90^\circ$, a $\mathrm{MR}_{\parallel}=+36\%$ for $\mu_{0}H_{\parallel}=$\,7\,T at $T=5\,\mathrm{K}$ is found, as shown in Fig.~\ref{fig:fig7}(e). Similar behavior is also observed for sample S2, as reported in Fig.~\ref{fig:fig7}(f). However, the reduced magnitude of $\mathrm{MR}_{\parallel}$ to -5\% for $\psi=0^\circ$ indicates that for this flake, $w$ is not aligned along the $b-$axis. The estimated values of $\mathrm{MR}_\perp$, $\mathrm{MR}_\parallel$ and $\mu_\mathrm{av}$ at $T=5\,\mathrm{K}$ for samples S1 and S2 are summarized in Table I. Additionally, two test samples are also studied, in which the metal contacts are fabricated on the exfoliated flakes by employing electron beam lithography (EBL). For both  samples, the measured resistance of the flakes is found to be $\sim$$10^5\,\Omega$ and the samples display electronic properties befitting of a semiconductor. Such a change in the electronic properties in samples with contacts fabricated on flakes $via$ EBL in comparison to the ones where the flakes are transferred onto the contacts, indicate that the exposure to chemicals and electron beams is detrimental for the semimetallic WTe$_{2}$ flakes considered in this work. Moreover, both Au and Pt are found to provide robust Ohmic contacts to thin exfoliated WTe$_{2}$ flakes.

\begin{figure*}[htbp]
	\centering
	\includegraphics[scale=1.75]{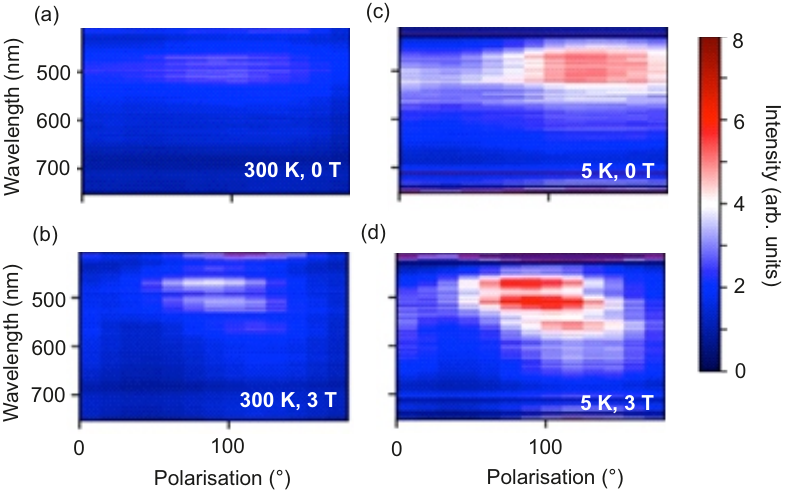}
	\caption{\label{fig:OR}  Reflectivity spectra for different $T$ and $\mu_{0}H$ as a function of the polarization for (a) $T=300\,\mathrm{K}$, $\mu_{0}H=0\,\mathrm{T}$; (b) $T=300\,\mathrm{K}$, $\mu_{0}H=3\,\mathrm{T}$; (c)  $T=50\,\mathrm{K}$, $\mu_{0}H=0\,\mathrm{T}$ and (d) $T=5\,\mathrm{K}$, $\mu_{0}H=3\,\mathrm{T}$.}	
	\label{fig:fig9}
\end{figure*}

\subsection*{Static optical reflectivity}

Static reflectivity measurements as a function of $T$, $\mu_{0}H_\perp$ and polarization of the incident light $P_\mathrm{L}$ have been performed on the exfoliated flakes. Optical reﬂectivity measured on WTe2 in the far infra-red (IR) and in the mid IR energy range was reported to point at charge compensation from electron and hole pockets at the Fermi level \cite{Homes:2015_PRB,Kimura:2019_PRB,Santos-Cottin:2020_PRMater}. Theoretical calculations based on $ab\,initio$ density functional theory have predicted anisotropic optical reflectivity in WTe$_{2}$ for an energy range $0.5\,\mathrm{eV}\leq{\hbar\omega}\leq3.5\,\mathrm{eV}$ as a function of the polarization \cite{Rano:2020_ResPhys}. Here, the experiments are carried out using a wide spectral range ultrafast pump-probe magneto-optical spectrometer at low-$T$/high-$\mu_{0}H$ \cite{Mertens:2020_RSI}. In particular, the probe beam of this set-up is sent through an additional element for the generation of supercontinuum pulses to perform reflectivity measurements. The details of the experimental set-up are provided in the Supplementary Material. The reflectivity spectra are recorded on a $\sim$45 nm thick exfoliated WTe$_{2}$ flake transferred onto a rigid SiO$_{2}$/Si$^{++}$ substrate provided with markers to facilitate the identification of the flake under a microscope. The spectra are collected in the wavelength range between 450 nm and 700 nm at $T = 5\,\mathrm{K}$ and at $T=300\,\mathrm{K}$ for $\mu_{0}H_\perp=0\,\mathrm{T}$ and for $\mu_{0}H_\perp=3\,\mathrm{T}$ and as a function of $P_\mathrm{L}$, where the linear polarization of the supercontinuum pulses is rotated within the sample plane.

The spectrum of the generated supercontinuum pulses, measured after passing the beam through a color filter and a polarizer, is shown in Fig.~\ref{fig:fig8}(a) for two perpendicular linear polarization directions corresponding to the electric field of light $E_\mathrm{L}$, defined as: (i) $P_\mathrm{L}=120^\circ\equiv{\parallel}$ and (ii)$P_\mathrm{L}=30^\circ\equiv{\perp}$, such that $E_\mathrm{L}$ is respectively parallel and perpendicular to the long axis $l$ of the measured WTe$_{2}$ flake, as shown in Fig.~\ref{fig:fig8}(b). The spectrum spans the spectral range between 450 nm to 700 nm and does not depend on the direction of the linear polarization. The measured reflectivity spectra are normalized with respect to a reference spectrum recorded by diverting the laser beam before the sample and are reported in Figs.~\ref{fig:fig9}(a)-(d) as a function of the polarization angle, of $T$, and of $\mu_{0}H_\perp$. The spectra are shown as 2-dimensional (2D) maps, where the intensity is encoded in color scale. As discussed in detail in the Supplementary Material, the reflectivity spectra recorded both at $T=300\,\mathrm{K}$ and $T=5\,\mathrm{K}$ show a dependence on $P_\mathrm{L}$, that is more pronounced for $\mu_{0}H_\perp=3\,\mathrm{T}$ than for $\mu_{0}H_\perp=0\,\mathrm{T}$. This anisotropic behavior is in agreement with what observed in the magnetotransport studies. Moreover, a comparison of the spectra recorded at a fixed $T$ reveals that the changes in the spectra after application of $\mu_{0}H_\perp=3\,\mathrm{T}$ are more pronounced for $P_\mathrm{L}=\parallel$ then for $P_\mathrm{L}=\perp$. This behavior is elucidated quantitatively by evaluating the anisotropy $A$ defined as:

\begin{equation}
A=\frac{R(3\,\mathrm{T})-R(0\,\mathrm{T})}{R(0\,\mathrm{T})}
\end{equation}

where $R(3\,\mathrm{T})$ and $R(0\,\mathrm{T})$ are the reflectivities measured for fixed light polarization and constant $T$ in the presence and absence of magnetic field, respectively. As expected, the asymmetry is significantly pronounced in the data recorded for parallel light polarization. Thus, the magneto-optical reflectivity measurements complement the magnetotransport results and support the existence of anisotropic features in the electronic structure responsible for the observed out-of-plane electronic properties of the thin $T_\mathrm{d}$-WTe$_{2}$ flakes.

\section*{Conclusion}

In summary, $T_\mathrm{d}$-WTe$_{2}$  flakes of thickness $\sim$45 nm have been fabricated $via$ mechanical exfoliation and transferred onto rigid SiO$_{2}$/Si$^{++}$ substrates with prepatterned electrical contacts using a viscoelastic dry transfer technique. The flakes are found to be stable in air and no chemical degradation is observed over an aging period of seven days. The two exemplary samples reported -- S1 and S2 -- are fabricated using 10 nm thick Pt metal pads as Ohmic contacts in a van der Pauw geometry and designed by photolithography. In S1, the relative orientation of the $T_\mathrm{d}$-WTe$_{2}$ flake w.r.t. the four contact pads leads to an exact alignment of the directions of the applied electric field $E$ and of the flake width $w$. In S2 the relative positions of the flake and the four Pt contacts result in a slight misalignment between $E$ and $w$. Raman spectroscopic measurements carried out at room temperature reveal five Raman active modes, matching the theoretical predictions. The samples exhibit a large $\mathrm{MR}_{\perp}$ as high as 1200\% at $T=5\,\mathrm{K}$ and for a $\mu_{0}H_\perp=7\,\mathrm{T}$ applied along the $c-$axis. A $\mu_{0}H_\perp$-dependent turn-on $T_\mathrm{Trans}$ is observed, below which the samples undergo a MIT originating from the anisotropy of the Fermi surface. Both samples follow a Fermi \mbox{liquid} behavior for $T\leq50\,\mathrm{K}$. The anisotropy of the Fermi surface -- in combination with the presence of electron and hole pockets in the electronic band structure leading to charge compensation -- is concluded to be at the origin of the large positive $\mathrm{MR}_\perp$. The calculated $\mu_\mathrm{av}$ $\sim$$5000$ cm$^{2}$V$^{-1}$s$^{-1}$ at $T=5\,\mathrm{K}$ for S1 is a property of the Fermi liquid, while for $T\geq50\,\mathrm{K}$ the carrier mobility monotonically decreases due to the dominant electron-phonon coupling. The observed negative $\mathrm{MR}_\parallel$ for $\mu_{0}H_{\parallel}\parallel\left({E}\parallel{w}\right)$ is a signature of  chiral anomaly in $T_\mathrm{d}$-WTe$_{2}$ and is found to be remarkably sensitive to the relative orientation of the $a-$ and $b-$axes w.r.t. the applied fields $\mu_{0}H_\parallel$ and $E$. The anisotropic behavior of the studied WSM-II system is confirmed by studying the optical reflectivity of the flakes as a function of $T$, $\mu_{0}H$ and polarization of $E_\mathrm{L}$ in the visible range of the electromagnetic sprctrum. It is also concluded, that the Weyl semimetallic properties of exfoliated thin flakes of WTe$_{2}$ are best observed when the flakes are transferred onto prefabricated metal Ohmic contacts, rather than when contacts are processed onto the flakes $via$ EBL. The tunability of the large positive $\mathrm{MR}_\perp$ and the chiral anomaly-driven negative $\mathrm{MR}_\parallel$ as a function of the crystal axes and thickness, in combination with the chemical stability, pave the way for the application of 2D WSM-II WTe$_{2}$ in the future generation of chiral electronic devices like $e.g.$ chiral batteries, and as active elements for the detection of ultraweak magnetic fields \cite{Kharzeev:2013_PRB}.

\section*{Acknowledgements}

The work was funded by the Austrian Science Fund (FWF) through Projects No. P26830 and No. P31423 and by the Deutsche Forschungsgemeinschaft (DFG) through the International Collaborative Research Center TRR142 (Project A08) and TRR160 (Projects B8 and B9). The authors thank Sabine Hild for discussions on the Raman spectroscopy data.

\bibliographystyle{apsrev4-2}

\end{document}